\documentclass[prd,aps,preprint,onecolumn,superscriptaddress,tightenlines,nofootinbib,eqsecnum,preprintnumbers,longbibliography,12pt]{revtex4-1}
\pdfoutput=1
\usepackage{amsmath,latexsym,amssymb,graphicx,ifpdf,slashed,color,hyperref,url,cancel}
\usepackage[T1]{fontenc}
\usepackage{relsize}
\hypersetup{colorlinks,citecolor= red ,linkcolor= niceblue, urlcolor=niceblue}
\definecolor{niceblue}{rgb}{0.1,0.2,0.6}

\begin{document}
\def\Carleton{Department of Physics, Carleton University, Ottawa, ON K1S 5B6, Canada}
\def\Loyola{Department of Physics, Loyola University Chicago, Chicago, IL 60660, USA}
\def\MPIK{Max-Planck-Institut für Kernphysik, Saupfercheckweg 1, 69117 Heidelberg, Germany}

\preprint{}

\title{Core-collapse Supernova Constraint on the Origin of Sterile Neutrino Dark Matter via Neutrino Self-interactions}

\author{Yu-Ming Chen}
\email{yumingchen@cmail.carleton.ca}
\affiliation{\Carleton}
\author{Manibrata Sen}
\email{manibrata@mpi-hd.mpg.de}
\affiliation{\MPIK}
\author{Walter~Tangarife}
\email{wtangarife@luc.edu}
\affiliation{\Loyola}
\author{Douglas Tuckler}
\email{dtuckler@physics.carleton.ca}
\affiliation{\Carleton}
\author{Yue Zhang}
\email{yzhang@physics.carleton.ca}
\affiliation{\Carleton}

\date{\today}
\begin{abstract}

Novel neutrino self-interaction can open up viable parameter space for the relic abundance of sterile-neutrino dark matter (S$\nu$DM). In this work, we constrain the relic target using core-collapse supernova which features the same fundamental process and a similar environment to the early universe era when S$\nu$DM is dominantly produced. We present a detailed calculation of the effects of a massive scalar mediated neutrino self-interaction on the supernova cooling rate, including the derivation of the thermal potential in the presence of non-zero chemical potentials from plasma species. Our results demonstrate that the supernova cooling argument can cover the neutrino self-interaction parameter space that complements terrestrial and cosmological probes.

\end{abstract}

\maketitle
\tableofcontents

\newpage
\section{Introduction}
Self-interaction among active neutrinos is a blind spot among the tests of the Standard Model (SM) of elementary particles. Laboratory experiments are yet to measure neutrino-neutrino scattering, due to their elusive nature. Although LEP has measured the neutrino coupling with the $Z$-boson \cite{ALEPH:2005ab}, the result does not exclude the possibility of stronger neutrino self-interaction through additional light mediators. Indeed, a viable wide window is open if the mediator is a SM gauge singlet that couples exclusively to neutrinos at tree level, known as a lepton number charged scalar~\cite{Berryman:2018ogk}, including the Majoron~\cite{Gelmini:1980re}. A neutrinophilic mediator can be probed in various ways, from neutrino experiments to cosmic microwave background (CMB) and stellar physics (see \cite{Berryman:2022hds} for a recent overview). The presence of such a new force can be strongly motivated as solutions to other significant problems, notably the origin and nature of dark matter that fills our universe.

The intimate relationship between neutrino self-interaction and sterile neutrino dark matter (${\rm S\nu DM}$) has been pointed out and explored in several recent works~\cite{DeGouvea:2019wpf, Kelly:2020pcy, Kelly:2020aks, Chichiri:2021wvw, Benso:2021hhh, Alonso-Alvarez:2021pgy}. Remarkably, with a sufficiently strong neutrino self-interaction, the ${\rm S\nu DM}$ relic target can evade the exclusion limits set by indirect X-ray searches~\cite{DeGouvea:2019wpf}. In the early universe, the self-interaction can keep the active neutrinos in thermal equilibrium among themselves for longer than the weak interaction allows, thereby increasing the active-sterile neutrino conversion rate. A complete portrait of the ${\rm S\nu DM}$ relic-density target parameter space was presented in Ref.~\cite{Kelly:2020aks}. It features a neutrinophilic mediator mass from keV to multi-GeV scale, and the corresponding mediator-neutrino coupling ranges from $\sim 10^{-8}$ to order 1. The viable mass of ${\rm S\nu DM}$ lies between keV and MeV. Compared to other ways of addressing the relic density problem, such a mechanism resorts to simple low-scale physics, which makes it highly testable. Phenomenologically, beam neutrino experiments like DUNE and the forward physics facility at the LHC can be useful for probing the GeV-scale mediator mass region of the relic target~\cite{Kelly:2019wow, Kelly:2021mcd, deGouvea:2019qaz}. In constrast, cosmological probes with precision $\Delta N_{\rm eff}$ measurement can cover the region with a very light, sub-MeV scale mediator~\cite{EscuderoAbenza:2020cmq, Kelly:2020aks, Chacko:2003dt}. However, an intermediate-mass window for the mediator is left open, where the coupling is too small to be relevant for terrestrial experiments, and the mediator is sufficiently heavy, thus evading the constraints from CMB and BBN. This part of the parameter space can be susceptible to cooling bounds arising out of core-collapse supernovae (CCSN). The present work's task is to fill such a gap by presenting a new supernova constraint on the relic target.

Our starting point is simple and natural. In the early Universe, ${\rm S\nu DM}$ is dominantly produced through secret neutrino self-interactions around a temperature of $\sim 1-100$\,MeV. A CCSN is another natural laboratory with a dense neutrino background where such temperatures can be reached in the core. As a result, ${\rm S\nu DM}$ can also be efficiently produced in the core of a CCSN. Production and emission of such non-interacting particles can result in excessive energy loss of the supernova (SN), leading to additional cooling. Observations of the neutrinos from SN1987A can constrain the amount of energy allowed to be carried away by any new and weakly-coupled particles~\cite{Raffelt:1996wa}. Stated simply, the neutrino burst must carry sufficient energy from the collapsed core to ignite the eventual explosion~\cite{Janka:2006fh} and last long enough to match the observation period of $\mathcal{O}(10)\,$s~\cite{Kamiokande-II:1987idp, Bionta:1987qt, 1987ESOC...26..237A}. This leads to an exclusion window in the parameter space of new light degrees of freedom, where they cool the SN fast, causing a shorter duration of neutrino burst. In general, the SN constraint can be evaded if the particles are too weakly coupled to be efficiently produced in
the core. On the other hand, if they are too strongly coupled, they will be reabsorbed back to the core, thereby not transporting the energy efficiently.

Although the fundamental process for ${\rm S\nu DM}$ production is identical between SN and early universe, there are important differences in the two analysis. While the early Universe is largely isotropic and homogeneous, a SN retains a lot less symmetry. The thermal plasma inside a SN is composed of  neutrons, protons, electrons and neutrinos, which have large matter-antimatter asymmetries. As a result, the forward scattering potential experienced by neutrinos picks up a non-trivial contribution from finite density effects, in addition to finite temperature contribution. Moreover, given the dense and fast-varying matter background, we need to consider the absorption and non-adiabatic effects on ${\rm S\nu DM}$ en route out of the SN. We expand on these points in details in the main text.

Excessive energy loss from a  CCSN has been considered before in the context of the minimal ${\rm S\nu DM}$ model without secret neutrino self-interactions. In the minimal ${\rm S\nu DM}$ model, the cooling constraint on the vacuum mixing angle is found less competitive than the existing X-ray search limit due to feedback effects~\cite{Suliga:2020vpz, Suliga:2019bsq, Arguelles:2016uwb, Raffelt:2011nc, Hidaka:2006sg, Dolgov:2000pj, Dolgov:2000jw}. On the other hand, the neutrinophilic mediator can be directly produced in the SN, leading to excessive cooling, as has been shown for tiny couplings~\cite{Escudero:2020ped, Brune:2018sab, Heurtier:2016otg, Farzan:2002wx, Kolb:1987qy}. Larger couplings of neutrino self-interactions can be constrained from the down-scattering in energy of SN neutrinos by their interaction with the cosmic neutrino background~\cite{Kolb:1987qy,Shalgar:2019rqe}. Interactions with a neutrinophilic scalar can also cause distinct changes in the SN collapse dynamics~\cite{Fuller-Mayle,Fuller:1988ega,Chang:2022aas}.
Constraints can also arise from the decay of the mediating boson to neutrinos, thereby altering the expected SN flux~\cite{Akita:2022etk}.
However, these constraints are not able to probe the ${\rm S\nu DM}$ relic target completely. In contrast, our work puts together  production of ${\rm S\nu DM}$ via neutrino self-interaction, in the SN core, and the excessive cooling due to free-streaming of ${\rm S\nu DM}$. We show that CCSN allows to explore new regions in the model's parameter space. This result is highly complementary to neutrino experimental and cosmological probes of the relic target.

The structure of this article is as follows. In Sec.~\ref{sec:formalism}, we write down the model and the formalism for calculating the SN cooling luminosity via ${\rm S\nu DM}$ emission. In particular, we derive the thermal potential generated by neutrino self-interaction with a massive mediator and non-zero chemical potentials. Next, we present our main result in Sec.~\ref{sec:results}, where we derive a new SN cooling constraint that can cover part of the dark matter relic target curve. We discuss the physics behind each edge of the exclusion region and highlight features unique to the cooling mechanism under consideration. We also present an overview of the model parameter space that shows where our main result stands along with other experimental probes of the origin of ${\rm S\nu DM}$. Our conclusions are placed in Sec.~\ref{sec:conclusion}.

\section{The formalism}
\label{sec:formalism}

\subsection{Model}

The model we consider involves two ingredients beyond the SM. First, sterile neutrino dark matter (${\rm S\nu DM}$), defined as a linear combination of active neutrinos $\nu$ (weak eigenstate) and a SM gauge singlet fermion $\nu_s$,
\begin{equation}
\nu_4 = \nu_s \cos\theta + \nu \sin\theta \ ,
\end{equation}
where $\nu_4$ is the mass eigenstate and $\theta$ is the active-sterile neutrino mixing angle in vacuum. Given the indirect-detection constraint on $\nu_4\to \nu\gamma$ decay, the angle $\theta$ needs to be very tiny. As a result, the above mixing plays no role in generating the observed neutrino masses, which need to be generated by other physics and can be either the Majorana or the Dirac type~\cite{Kelly:2020pcy}.

The second ingredient is new neutrino self-interactions mediated by a massive scalar boson $\phi$. The relevant interacting Lagrangian takes the form
\begin{equation}\label{eq:nuselfint}
\mathcal{L}_{\rm int} = \frac{(L_\alpha H)(L_\beta H) \phi}{\Lambda_{\alpha\beta}^2} + {\rm h.c.} \to \frac{1}{2} \lambda_{\alpha\beta} \nu_\alpha\nu_\beta\phi  + {\rm h.c.} \ .
\end{equation}
The effective operator in the first step is SM gauge invariant. In the second step, we turn on the electroweak symmetry breaking where the Higgs vacuum expectation value makes the $\phi$ coupling neutrinophilic. The flavor indices $\alpha,\beta = e, \mu, \tau$ are summed over.

\begin{figure}[t]
	\centerline{\includegraphics[width=0.5\textwidth]{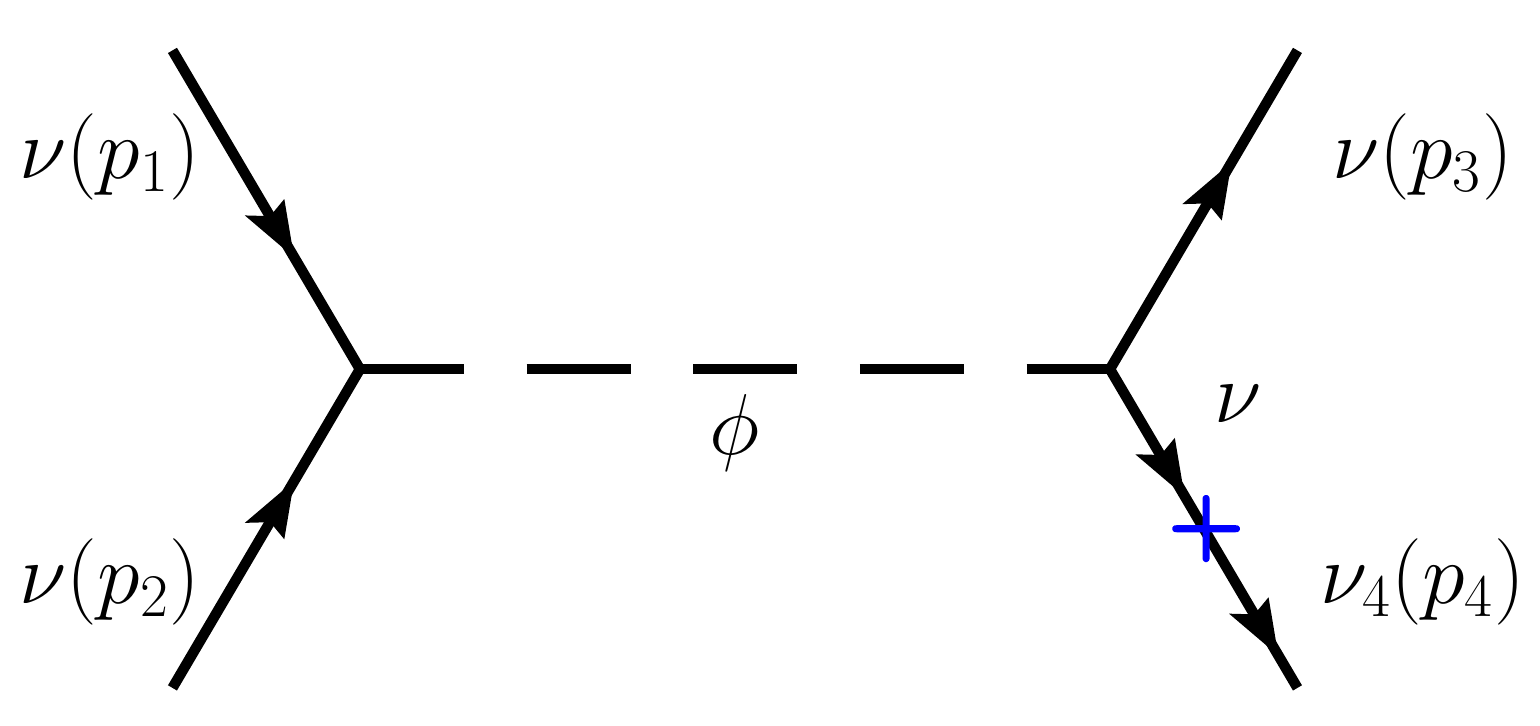}}
	\caption{Identical fundamental process governs the production of ${\rm S\nu DM}$ in early universe (for relic density) and the excessive cooling of CCSN.
	The blue cross represents the effective active-sterile neutrino mixing in the corresponding epoch. Time flows from left to right.}\label{fig:FD}
\end{figure}

Within this simple framework, the correct relic abundance for ${\rm S\nu DM}$ can be produced via active-sterile neutrino oscillation in the early universe~\cite{DeGouvea:2019wpf}. In the early universe, the self-interaction can keep the active neutrinos in thermal equilibrium with themselves. After the production of
an active neutrino from the thermal plasma, it oscillates into the sterile
neutrino state until another neutrino self-interaction occurs. Because of oscillation, there is a small probability for the coherent state to be measured as the dark matter $\nu_4$. A Feynman diagram summarizing the above processes is shown in Fig.~\ref{fig:FD}. The same story applies to all neutrinos in the thermal plasma and can repeat many times until the self-interaction decouples, leading to a buildup of the ${\rm S\nu DM}$ abundance. Because the $\phi$-mediated neutrino self-interaction is allowed to be much stronger than the weak self-interactions, the correct relic abundance can be produced with a sufficiently small mixing angle $\theta$ that satisfies the indirect detection X-ray constraint.

\subsection{Active-Sterile Neutrino Mixing in Supernova}

The active-sterile neutrino effective mixing angle inside a supernova (SN) is calculated in a similar way as its early universe counterpart~\cite{Dodelson:1993je},
\begin{equation}\label{eq:theta_eff}
\sin^2 2\theta_{\rm eff} (E, r) = \frac{\Delta^2 \sin^2 2\theta}{\Delta^2 \sin^2 2\theta + \Gamma (E, r)^2
+ (\Delta \cos2\theta - V(E, r))^2} \ ,
\end{equation}
where $\Delta \simeq m_4^2/(2E)$, and $\Gamma = \Gamma_{\rm weak} + \Gamma_\phi$ and $V = V_{\rm weak} + V_\phi$ are the interaction rate and effective potential seen by an active neutrino (or antineutrino), respectively.  Because SN features a sizable matter--antimatter asymmetry, neutrino and antineutrino typically see different $\Gamma$ and $V$ and, in turn, very different effective mixing angles.

\begin{figure}[t]
	\centerline{\includegraphics[width=0.9\textwidth]{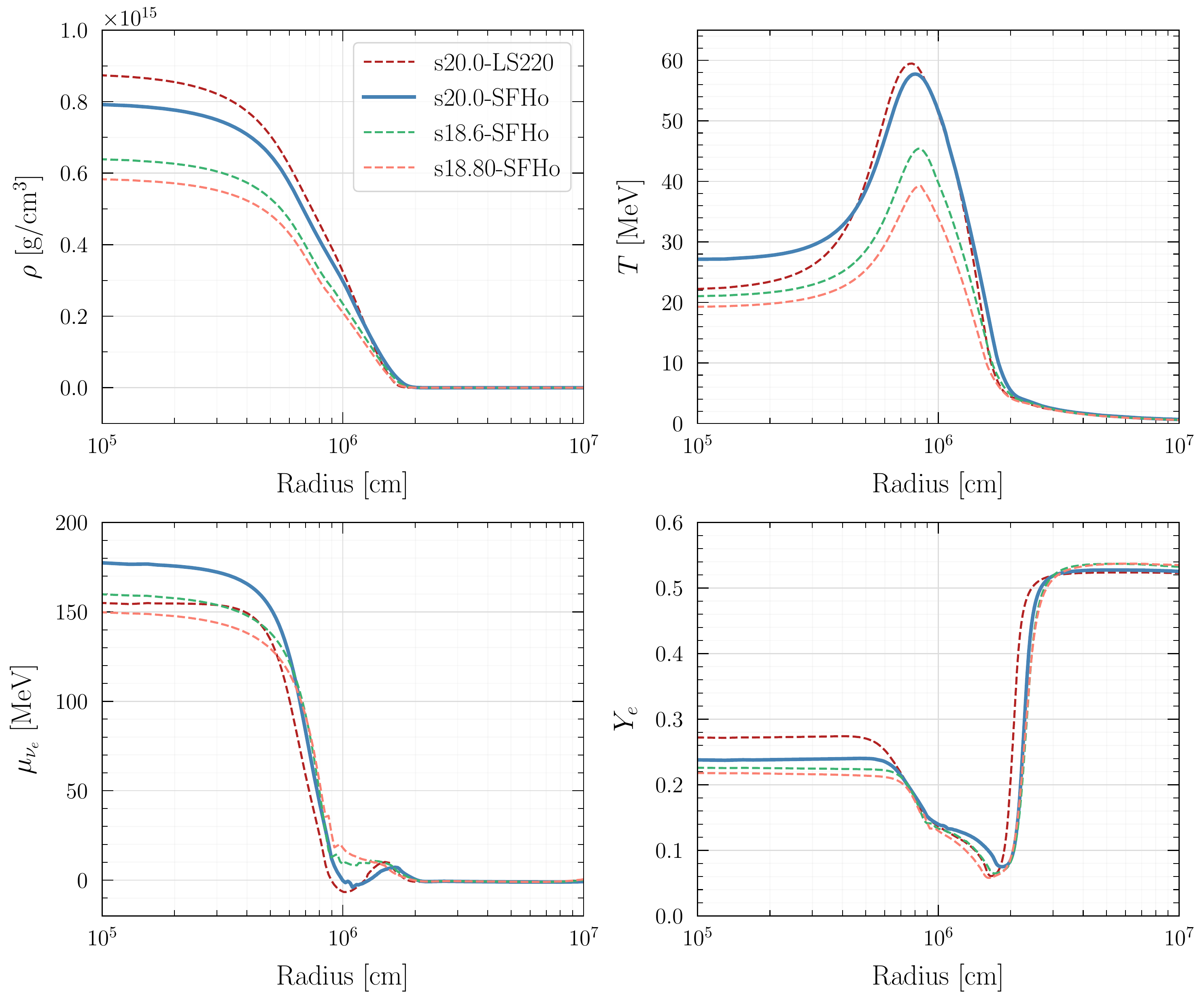}}
	\caption{Mass density $\rho$, temperature $T$, chemical potential $\mu_{\nu_e}$, and electron fraction $Y_e$ as functions of the radial coordinate $r$, taken from hydrodynamical simulations~\cite{GarchingGroup}. 
	The SN model we use is {\tt s20.0-SFHo}, as shown by the solid blue curves.
	For comparison, we also show a few other models (in colored dashed curves) used in the literature.
	}\label{fig:profiles}
\end{figure} 

The neutrino interaction rate through weak interactions $\Gamma_\text{weak}$ is dominated by its scattering with the ambient neutrons and protons via $Z$- and $W$-boson exchange. For neutrino energy $E$ well below the neutron mass,
\begin{equation}\label{eq:Gamma_weak}
\Gamma_\text{\rm weak}(E, r) = \frac{(1+3 g_A^2)\,G_F^2 E^2}{4\pi} \times
\left\{
\begin{array}{cl}
n_n(r), &\quad{\rm for}\ \nu_\mu, \bar\nu_\mu, \nu_\tau, \bar\nu_\tau \\
5 n_n(r), &\quad{\rm for}\ \nu_e \\
n_n(r)+4n_p(r), &\quad{\rm for}\ \bar\nu_e \\
\end{array}
\right.
\end{equation}
where $g_A\simeq1.3$ is the low-energy nucleon axial coupling, and $n_i(r)$ is the number density of particle $i$. Hereafter, we assume the SN is a spherical symmetric object. We also neglect the mass of electron and proton-neutron mass difference which are much smaller than the typical energy scales of the CCSN, e.g. core temperature and densities.

The neutrino interaction rate through $s$-channel $\phi$-exchange is~\cite{Kelly:2020aks}
\begin{equation}\label{eq:Gamma_phi}
\Gamma_\phi(E, r) = \frac{|\lambda|^2\, m_\phi^2\, T(r)}{16\pi\, E^2} \left[ \log\left( e^{\frac{\mu_\nu(r)}{T(r)}} + e^{\frac{m_\phi^2}{4 E T(r)}} \right) - \frac{m_\phi^2}{4 E \,T(r)} \right] \ ,
\end{equation}
where $T(r)$ is the matter temperature and $\mu_\nu(r)$ is the neutrino chemical potential inside the SN. The latter is nonzero for $\nu_e$ but neglected for $\nu_\mu$ and $\nu_\tau$. For the sake of simplicity, we have dropped the flavor indices of the coupling $\lambda$. In the following analysis, we will assume only one such coupling is around each time, i.e., the same active neutrino that mixes with $\nu_s$ also participates in the novel self-interaction introduced in Eq.~\eqref{eq:nuselfint}.

Because SN is made of asymmetric matter, the leading contribution to the matter potential through weak interactions occurs at order $\mathcal{O}(G_F)$~\cite{Abazajian:2001nj},
\begin{equation}\label{eq:V_weak}
V_{\rm weak}(E, r) = \left\{
\begin{array}{cl}
\sqrt{2} G_F \left[ 2 (n_{\nu_e} - n_{\bar\nu_e}) + (n_{e^-} - n_{e^+}) -\frac{1}{2} (n_n-n_{\bar n})  \right], &\quad{\rm for}\ \nu_e \\
\sqrt{2} G_F \left[ (n_{\nu_e} - n_{\bar\nu_e}) -\frac{1}{2} (n_n-n_{\bar n})  \right], &\quad{\rm for}\ \nu_{\mu, \tau} \\
\end{array}
\right.
\end{equation}
where the number densities depend on $r$.
The potentials for antineutrino have opposite sign.
The temperature-dependent terms occur at order $\mathcal{O}(G_F^2)$ and are negligible. It is useful to note that $V_{\rm weak}$ depends on a lower power of $G_F$ than $\Gamma_{\rm weak}$, which leads to an interesting interplay when neutrino self-interaction is turned on as will be discussed in Sec.~\ref{sec:results}.

We employ the SN profile titled {\tt s20.0-SFHo} from hydrodynamical simulations by the Garching  group~\cite{GarchingGroup}. Fig.~\ref{fig:profiles} shows the $r$ dependence of temperature $T$, chemical potential for the electron-type neutrino $\mu_{\nu_e}$, mass density $\rho$, and electron fraction $Y_e$. Our calculations are based on the profiles at time $t=1\,$s., which is characteristic of the cooling phase of the neutrino emission. 
For a given $r$, the particle asymmetries in neutron and electron are related to the electron fraction, $Y_e$, as
\begin{equation}
n_n - n_{\bar n} = \frac{\rho}{m_n} (1- Y_e) \ , \quad\quad
n_{e^-} - n_{e^+} = \frac{\rho}{m_n} Y_e \ .
\end{equation}
On the other hand, the particle asymmetry in ultra-relativistic neutrino is derived assuming a thermal distribution,
\begin{equation}
n_{\nu_e} - n_{\bar \nu_e} = - \frac{3T^3}{\pi^2} \left[ {\rm Li}_3 \left( - e^{\mu_{\nu_e}/T} \right) - {\rm Li}_3 \left( - e^{- \mu_{\nu_e}/T} \right) \right] \ ,
\end{equation}
where ${\rm Li}_3$ is the third-order polylogarithm function.

\begin{figure}[t]
	\centerline{\includegraphics[width=0.5\textwidth]{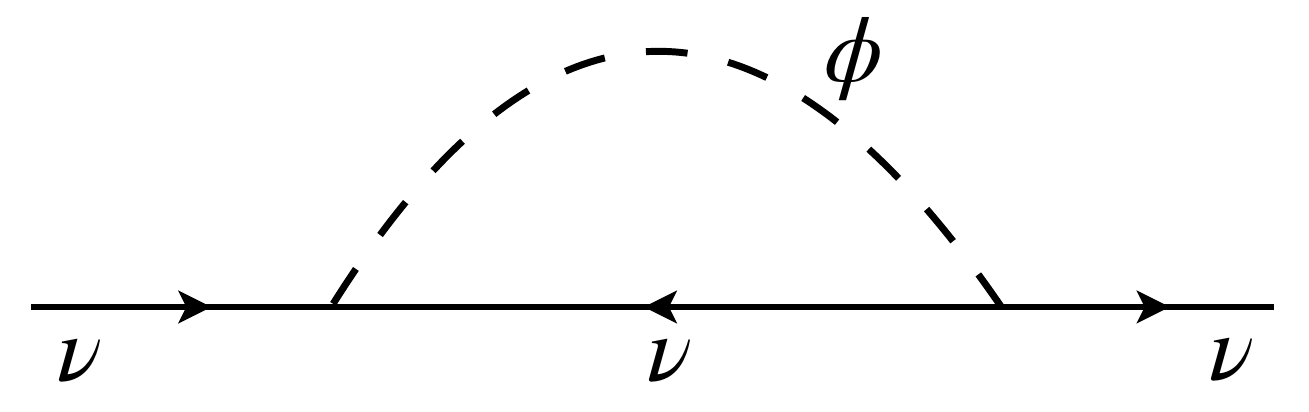}}
	\caption{Self-interaction contribution to the self-energy potential, $V$, in Eq.~\eqref{eq:V_phi}.}\label{fig:SE}
\end{figure}

Last but not least, we need $V_\phi$, the thermal potential generated by neutrino self-interaction.
In the past, the integral form of self-interaction contribution to the potential $V_\phi$ was derived for a generic mediator mass but assuming vanishing chemical potential in Ref.~\cite{Quimbay:1995jn}. In the presence of a chemical potential, the potential was derived assuming a massless mediator in Ref.~\cite{Erdas:1993qj}. In this work, we derive the most general form of the thermal potential for a neutrino with energy $E$, taking into account both the chemical-potential and mediator-mass effects. We find that the potential arising from the neutrino self-energy diagram, as shown in Fig.~\ref{fig:SE}, is given by
\begin{equation}\label{eq:V_phi}
\begin{split}
V_{\phi}(E, r) &= \frac{|\lambda|^2}{16\pi^2} {\rm Re}\int_0^\infty dk \left[ \left( \frac{m_\phi^2}{2E^2} \log\frac{m_\phi^2}{m_\phi^2 + 4 E k} + \frac{2k}{E} \right) \frac{1}{e^{(k+\mu_\nu(r))/T(r)}+1} \right. \\
&\hspace{3cm} \left. + \left( - \frac{m_\phi^2}{2E^2} \log\frac{m_\phi^2}{m_\phi^2 - 4 E k} + \frac{2k}{E} \right) \frac{1}{e^{(k-\mu_\nu(r))/T(r)}+1} \right] \\
&+ \frac{|\lambda|^2}{16\pi^2} {\rm Re}\int_0^\infty 
\frac{
kdk
}{
\epsilon_1
}
\left[
\left( \frac{2k}{E} -  \frac{m_\phi^2}{2E^2} \log \frac{m_\phi^2 + 2 E k + 2 E \epsilon_1}{m_\phi^2 - 2 E k + 2 E \epsilon_1} \right)  \frac{1}{e^{(\epsilon_1-2\mu_\nu(r))/T(r)}-1} 
\right. \\
&\hspace{3cm}+ \left.\left( \frac{2k}{E} - \frac{m_\phi^2}{2E^2} \log \frac{m_\phi^2 + 2 E k - 2 E \epsilon_1}{m_\phi^2 - 2 E k - 2 E \epsilon_1}\right) \frac{1}{e^{(\epsilon_1+2\mu_\nu(r))/T(r)}-1} \right] \ ,
\end{split}
\end{equation}
where $\epsilon_1\equiv\sqrt{k^2+m_\phi^2}$. The first integral corresponds to cutting the neutrino propagator whereas the second one corresponds to cutting the $\phi$ propagator. For the thermal distribution of $\phi$, we have imposed the chemical potential relation $\mu_\phi = 2 \mu_\nu$. For the case of antineutrino, one simply needs to flip the sign of chemical potential, i.e., $\mu_\nu \to \mu_{\bar \nu}=-\mu_\nu$.

We provide the details of the derivation of Eq.~\eqref{eq:V_phi} in Appendix~\ref{app:0}. We have also verified that, in the $\mu\to 0$ or $m_\phi\to 0$ limits, $V_{\phi}(E, r)$ reduces to the results in Refs.~\cite{Quimbay:1995jn} or~\cite{Erdas:1993qj}, respectively.

\subsection{Collisional Cooling Rate}
\label{sec:collisionalcooling}

The new SN cooling process we consider is depicted in Fig.~\ref{fig:FD}, where two active neutrinos scatter, via $s$-channel, into an on-shell $\phi$ particle, which quickly decays into an active and a sterile neutrino. The decay of $\phi$ involves the same Yukawa coupling as its production and the active-sterile neutrino mixing angle inside the SN. The corresponding luminosity is defined as energy release in the form of S$\nu$DM $\nu_4$ per unit time,
\begin{equation} \label{eq:Lcool}
\begin{split}
L &= \int d^3 \vec{r} \int \frac{d^3 \vec{p}_1}{(2\pi)^3} f_\nu(E_1, r) \int \frac{d^3 \vec{p}_2}{(2\pi)^3} f_\nu(E_2, r) \frac{1}{4 E_1 E_2} \int \frac{d^3 \vec{p}_3}{(2\pi)^3 2E_3} \int \frac{d^3 \vec{p}_4}{(2\pi)^3 2E_4} \\
&\quad \times (2\pi)^4 \delta^4 (\vec{p}_1+\vec{p}_2 - \vec{p}_3 - \vec{p}_4)
|\mathcal{M}|^2 \, E_4\, e^{- \tau (E_4, r)}\ ,
\end{split}
\end{equation} 
where 
\begin{equation}
f_\nu(E, r) = \frac{1}{e^{\frac{E-\mu_\nu(r)}{T(r)}} + 1} \ 
\end{equation}
is the Fermi-Dirac distribution function for an initial state neutrino inside core-collapse supernova (CCSN). Both the chemical potential and temperature depend on the location where the production of $\nu_4$ occurs. In the last factor of Eq.~\eqref{eq:Lcool}, $\tau$ is the optical depth, which measures how likely it is for $\nu_4$ to re-scatter and lose energy back to the SN on its way out, and is defined as
\begin{equation}
\tau(E, r) = \int_r^{\infty} dr_1 \sin^2\theta_{\rm eff} (E, r_1) \Gamma (E, r_1)\ .
\end{equation}
A more accurate derivation of the optical depth involves taking into account the attenuation of $\nu_4$ produced from the source point $\vec{r}_1$ in all the angular directions~\cite{Caputo:2022rca}. However, as we will show below (see sec.~\ref{sec:results}), the re-absorption only has a subdominant effect in the cooling of SN. For simplicity, we use the above approximate form of the optical depth~\cite{Chang:2016ntp, Chang:2018rso}.

Under the narrow-width approximation, the squared matrix element is equal to
\begin{equation}\label{eq:M2}
|\mathcal{M}|^2 = 32 \pi^2 \lambda^2 \, m_\phi^2 \,\delta (s-m_\phi^2) \sin^2 \theta_{\rm eff}(r, E_4) \ ,
\end{equation}
where $\theta_{\rm eff}$ is the effective mixing angle between the active and sterile neutrinos, defined in Eq.~\eqref{eq:theta_eff}, and depends on the energy $E$ of the state and its position $r$.

Clearly, due to the extra factor of $E_4$, the cooling rate defined in Eq.~\eqref{eq:Lcool} is not Lorentz invariant. The physical cooling rate is defined in the rest frame of the CCSN. We first complete the final-state phase-space integrals in a reference frame with generic $\vec{p}_1 + \vec{p}_2$, and then simplify the initial-state integral using the $\delta$-function in $|\mathcal{M}|^2$ (Eq.~\ref{eq:M2}). See Appendix~\ref{app:A} for further details. The energy loss rate carried away by $\nu_4$ is finally reduced to
\begin{equation} \label{eq:Lcool2}
\begin{split}
L &= \frac{\lambda^2 m_\phi^2}{4\pi^2} \int_0^{4 R_c} r^2dr \int_0^\infty dE_1 f(E_1, r) \int_{m_\phi^2/(4E_1)}^\infty dE_2 f(E_2, r) \frac{1}{\sqrt{(E_1+E_2)^2 - m_\phi^2}} \\
&\quad \times \int_{\frac{1}{2}\left( E_1+E_2 - \sqrt{(E_1+E_2)^2 - m_\phi^2} \right)}^{\frac{1}{2}\left( E_1+E_2 + \sqrt{(E_1+E_2)^2 - m_\phi^2} \right)} dE_4 \sin^2 \theta_{\rm eff}(r, E_4)\, E_4\, e^{- \tau (E_4, r)} \ .
\end{split}
\end{equation}
The upper limit of $r$ integral is taken to be $4 R_c\,=\,40\,$km. The last three integrals are evaluated numerically. Our main results will be presented in Sec.~\ref{sec:results}.

\section{Results}\label{sec:results}

\subsection{New supernova cooling limit}

We consider three benchmark choices of the model parameters that are (marginally) allowed by the present $X$-ray limits~\cite{Kopp:2021jlk, Abazajian:2017tcc, Roach:2022lgo},
\begin{align}
    &{\rm BP1:} \quad m_4\, =\, 7\,\,{\rm keV}\,,  &&\sin^22\theta \, =\, 7\times10^{-11} \ , \nonumber\\
    &{\rm BP2:} \quad m_4 \, =\, 21\,\,{\rm keV}\,,  &&\sin^22\theta \, =\, 1.4\times10^{-13} \ ,\label{eq:bps} \\
    &{\rm BP3:} \quad m_4 \, =\, 50\,\, {\rm keV}\,,  &&\sin^22\theta \, =\, 10^{-14} \ . \nonumber
\end{align}
For each benchmark point, we consider two cases where the neutrino chemical potential is present in the SN (for $\nu_e$) and absent (for $\nu_\mu$ and $\nu_\tau$). As mentioned earlier, we assume the same active neutrino that mixes with $\nu_s$ also participates in the novel self-interaction introduced in Eq.~\eqref{eq:nuselfint}.
We assume one coupling for the self-interaction to be present in each analysis.

\begin{figure}[t]
    \centerline{
    \includegraphics[width=1.0\textwidth]{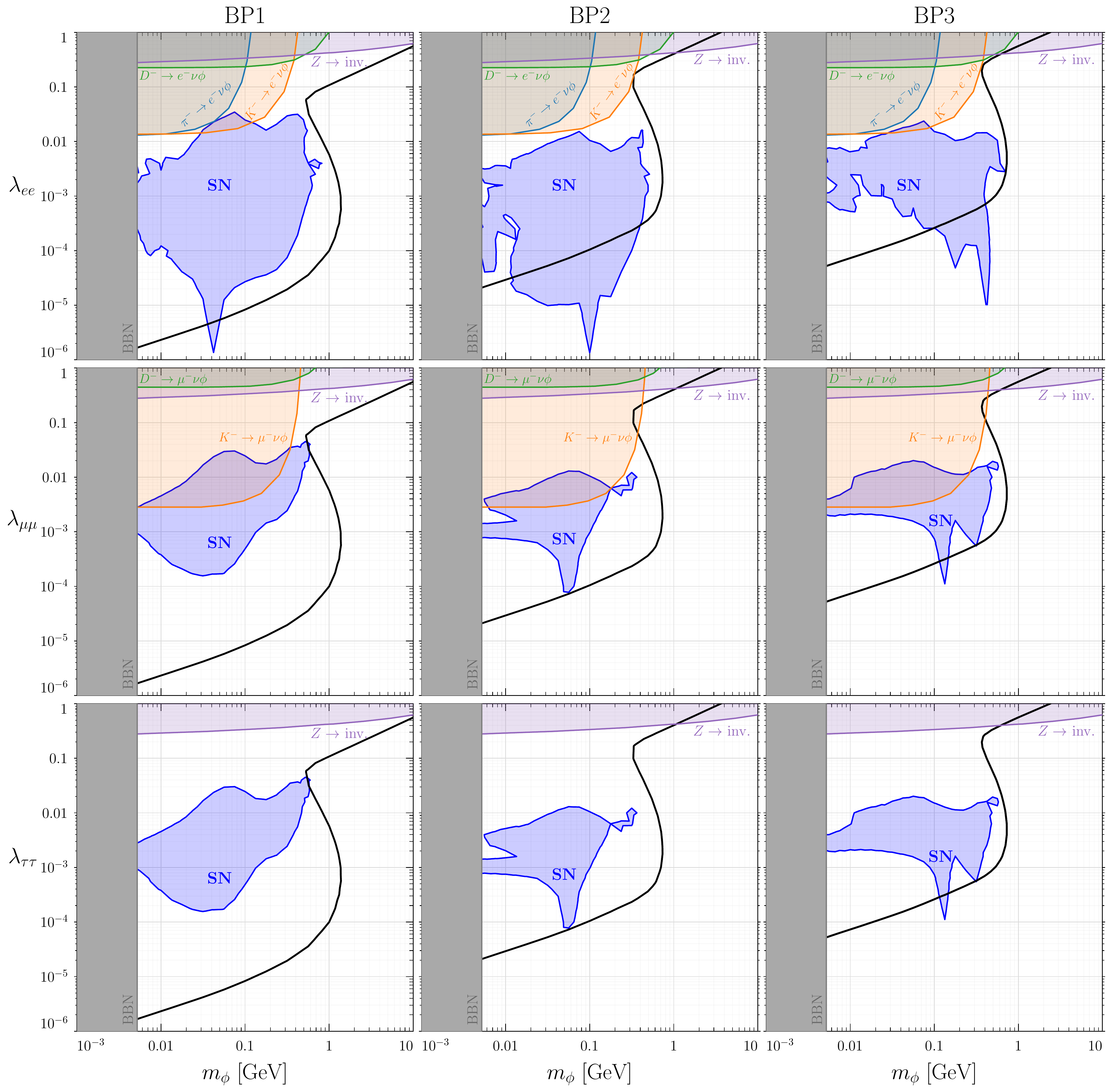}}
	\caption{Core-collapse supernova constraint derived in this work (blue shaded regions, labelled by ``SN'') on the parameter space of neutrino self-interaction that can address the relic density of S$\nu$DM (solid black curves). We consider three benchmark points defined in Eq.~\eqref{eq:bps} and three flavored coupling choices of the mediator $\phi$ that are labelled correspondingly. In each panel of the figure, we also show other constraints from precision measurements of the $Z$-boson, meson decays (purple, orange, green shaded regions), and big-bang nucleosynthesis (gray shaded regions).}\label{fig:mainresult}
\end{figure}

Our main result is shown in Fig.~\ref{fig:mainresult}, where we required that the luminosity of S$\nu$DM emission from SN-1987A must satisfy the upper bound\footnote{This is often referred to as the Raffelt bound~\cite{Raffelt:1996wa}. It is worth noting that when constraining any new-physics models, the most accurate determination of this upper bound would require incorporating the new physics in the original SN simulation. This has to be done point by point in the model parameter space, which we do not attempt here. Our goal is to work at the order-of-magnitude level and demonstrate the potential relevance of the SN cooling limit.}
\begin{equation}\label{eq:Raffelt}
L \lesssim 3\times10^{52}\,{\rm erg/s}\ .
\end{equation}
This excludes the blue shaded regions in the $\lambda$-$m_\phi$ plane. In each panel, the black curve is where the correct relic density for the S$\nu$DM ($\nu_4$) can be produced by neutrino self-interaction mediated by $\phi$, as suggested in~\cite{DeGouvea:2019wpf}. Our results show that observations from SN1987A can cover regions with relatively smaller coupling $\lambda$ compared to the laboratory limits and 
set a useful constraint on the origin of dark matter in this model.

For all three benchmark points, the SN constraint applies to $\phi$ masses up to several hundred MeV. Because the cooling channel considered in this work relies on the production of on-shell $\phi$ (that promptly decays into the $\nu_4$) and the typical temperature of the SN core is at most $\sim 60\,$MeV, the luminosity gets strongly Boltzmann suppressed for $\phi$ heavier than several hundred MeV and the cooling constraint goes away.

\begin{figure}[t]
    \centerline{\includegraphics[width=1.0\textwidth]{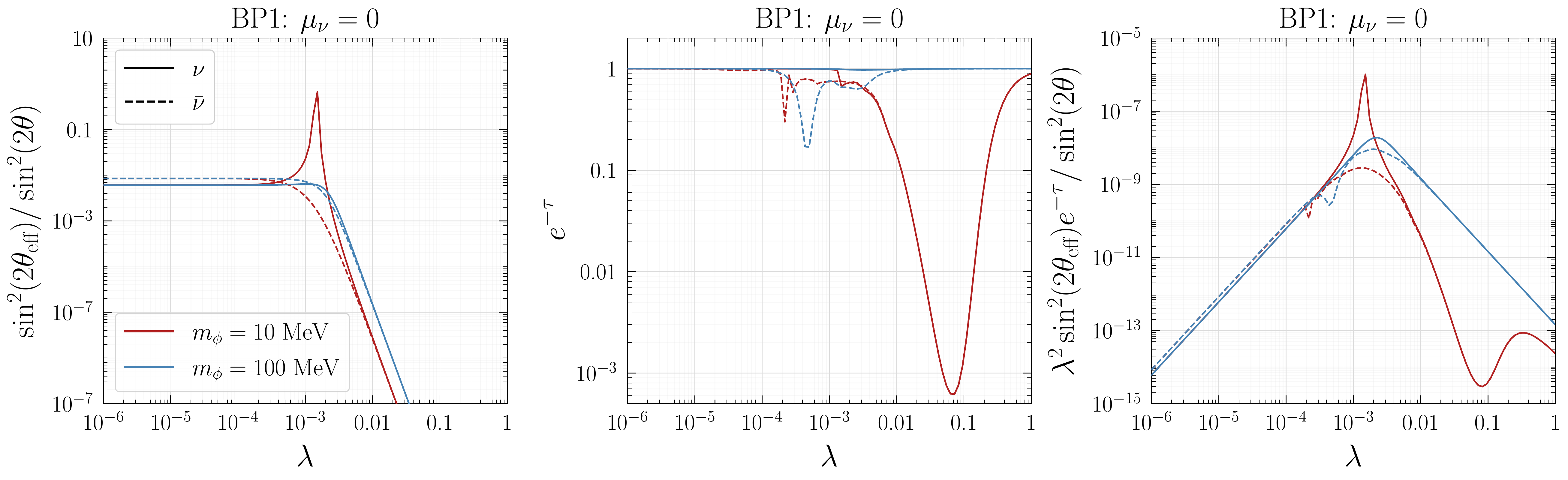}}
	\caption{The $\lambda$ dependence of key parameters in the SN cooling rate. From left to right: $\sin^22\theta_{\rm eff}/\sin^22\theta$, $e^{-\tau}$ and $\lambda^2 \sin^22\theta_{\rm eff} e^{-\tau}$. An important message here is that the effective mixing suppression plays the dominant role in large coupling region, $\lambda > 10^{-3}$. This effect applies generically to all mediator masses and overshadows the re-absorption of $\nu_4$. Here we show the plots for model point BP1 defined in Eq.~\eqref{eq:bps}. The corresponding plots are similar for BP2 and BP3, and when the chemical potential in switched on (for the $\nu_e$ case) .}\label{fig:anatomy}
\end{figure}

Regarding the Yukawa coupling $\lambda$ between $\phi$ and active neutrino $\nu$, the SN constraint is the strongest for $\lambda$ around $10^{-3}$. To understand this feature, we present Fig.~\ref{fig:anatomy}, which shows the $\lambda$ dependence of $\lambda^2 \sin^22\theta_{\rm eff} \,e^{-\tau}$, a key factor in the total luminosity calculation (see Eq.~\eqref{eq:Lcool2}). Here, for clarity, we fix the energy and production radius of the final state $\nu_4$ to the typical values for SN 1987A, $E_4 \sim T \sim 30\,$MeV and $r \sim R_c$.

\begin{itemize}

\item For very small values of $\lambda$, the neutrino self-interaction is too weak to modify the effective active-sterile neutrino mixing angle anywhere inside the SN, thus $\theta_{\rm eff} \sim \theta$. In this case, the energy loss rate $L$ is simply proportional $\lambda^2$. As $\lambda$ grows, the effective potential energy and scattering rate for neutrino generated by $\phi$ exchange exceed those due to the weak interaction, and the effective active-sterile neutrino mixing becomes strongly suppressed for larger $\lambda$. This is similar to the ``quantum Zeno'' effect that hampers the S$\nu$DM production in the very early universe~\cite{Dodelson:1993je}. The transition from ``small'' to ``large'' $\lambda$ regime occurs when the potential $V_\phi$, given in Eq.~\eqref{eq:V_phi}, is comparable to $V_{\rm weak}$ in Eq.~\eqref{eq:V_weak}~\footnote{Here, for a back-of-the-envelope estimate, we use the approximations, $V_{\rm weak} \sim G_F n_n$ and $V_\phi \sim \lambda^2 E_4/(16\pi^2)$},
\begin{equation}
\lambda \sim 4\pi \sqrt{G_F n_n/E_4} \sim 10^{-3} \ .
\end{equation}

\item In the large-$\lambda$ regime, the neutrino self-interaction rate due to $\phi$ exchange, $\Gamma_\phi \propto \lambda^2$, is also much higher than the weak interaction counterpart, $\Gamma_{\rm weak}$. As a result, the effective mixing is approximately $\sin^2\theta_{\rm eff} \sim \Delta^2\sin^2{\theta}/\Gamma_\phi^2 \sim \Delta^2\sin^2{\theta}/\lambda^4$, and in turn, the luminosity is proportional to $L \sim \lambda^2 \sin^2\theta_{\rm eff} \sim \lambda^{-2}$. The latter is thus suppressed for increasing $\lambda$. The above features are clearly displayed by the plots in the first column of Fig.~\ref{fig:anatomy}. The second column of Fig.~\ref{fig:anatomy} plots another factor in the cooling luminosity, $e^{-\tau}$, which describes the survival probability of $\nu_4$ against re-absorption on its way out of the SN. We find the absorption effect to be a subdominant effect. It only occurs in the large $\lambda$ regions where $\sin^22\theta_{\rm eff}$ is already strongly damped. The above anatomy reveals an important distinction between the SN cooling effect derived in this model and many others -- in the large coupling regime, the cooling rate is small mainly due to the dynamical suppression in the effective active-sterile neutrino mixing parameter rather than the absorption effect. Finally, the overall $\lambda$ dependence in the quantity $\lambda^2 \sin^22\theta_{\rm eff} e^{-\tau}$ is displayed by the plots in the third column of Fig.~\ref{fig:anatomy}.

\end{itemize}

Our result in Fig.~\ref{fig:mainresult} shows that the $\lambda_{ee}$ coupling receives a much stronger constraint from SN compared to $\lambda_{\mu\mu}$ and $\lambda_{\tau\tau}$. This is mainly because of the large chemical potential of $\nu_e$ in the SN (see Fig.~\ref{fig:profiles}). In the core region, the ratio $\mu_{\nu_e}/T$ is much larger than 1. As a result, the Fermi-Dirac distribution for $\nu_e$ remains not exponentially suppressed in the region $T < E < \mu_{\nu_e}$. Effectively, the large chemical potential creates a ``Fermi surface'' below which the occupation number of each energy state is order 1. The integral over the above energy window contributes and enhances the total cooling rate, Eq.~\eqref{eq:Lcool2}, and allows the SN coverage to expand significantly.

In Fig.~\ref{fig:mainresult}, we also depict other existing constraints on the model parameter space. They include precision measurements of the invisible $Z$-boson decay (new contribution from $Z\to\nu\nu\phi^*$), leptonic decay of charged pion, kaon and $D$ meson (new contribution channel $\mathfrak{m}^- \to \ell^- \nu \phi^*$)~\cite{Pasquini:2015fjv, Berryman:2018ogk, Blinov:2019gcj, Brdar:2020nbj, Lyu:2020lps, Esteban:2021tub}. In addition, we also include the constraint from big-bang nucleosynthesis which sets a lower limit on the mediator $\phi$ mass. For the range of coupling considered in the figure, $\phi$ is necessarily thermalized due to its interaction with active neutrinos before the weak interaction decoupling and can contribute to $\Delta N_{\rm eff}$ if its mass is around MeV or below.

We want to stress again that the new SN limits derived here rely on the presence of a light sterile neutrino with certain mixing with the active neutrinos, as defined in Eq.~\eqref{eq:bps}. With the mixing angle and proper neutrino self-interaction, it can obtain the correct relic density and serve as all the dark matter in the universe. This differs from any previous works and is exactly the main motivation and starting point of this work. Our result shows that observations from a CCSN can be useful for covering a new part on the S$\nu$DM relic target, especially for the case of $\lambda_{ee}$ coupling.

\subsection{Comment on Resonant Transition}

In this subsection, we comment on an alternative production channel of S$\nu$DM that could occur in the CCSN, via non-adiabatic transition between two energy levels due to the Mikheyev-Smirnov-Wolfenstein (MSW) resonance~\cite{Mikheev:1986if,Wolfenstein:1977ue}, and argue that such an effect is negligible in the presence of neutrino self-interactions. 

The MSW resonance condition can be satisfied if the active neutrino scattering is not very frequent. The effective active-sterile mixing angle defined in Eq.~\eqref{eq:theta_eff} can reach $\pi/4$ at a radius where $\Delta \cos2\theta = V(r)$ if the $\Gamma$ term in the denominator can be neglected. The latter requires the active neutrino to be free-streaming within a region where the resonance occurs. To quantify this, it is useful to introduce the spatial width of the resonance region, $\Delta_{\rm Res}$, and the corresponding neutrino free-streaming length, $\lambda_\nu$. They are defined by
\begin{align}
\Delta_{\rm Res} &\equiv 2 \tan(2\theta) \left| \frac{V'}{V} \right|^{-1} \ , \label{eq:Delta_res} \\
\lambda_\nu & \equiv 1/\Gamma \ . \label{eq:lambda_nu}
\end{align}
In Appendix~\ref{app:B}, we present a derivation showing why $\Delta_{\rm Res}$ is the relevant quantity. 

\begin{figure}[t]
    \centerline{\includegraphics[width=1.0\textwidth]{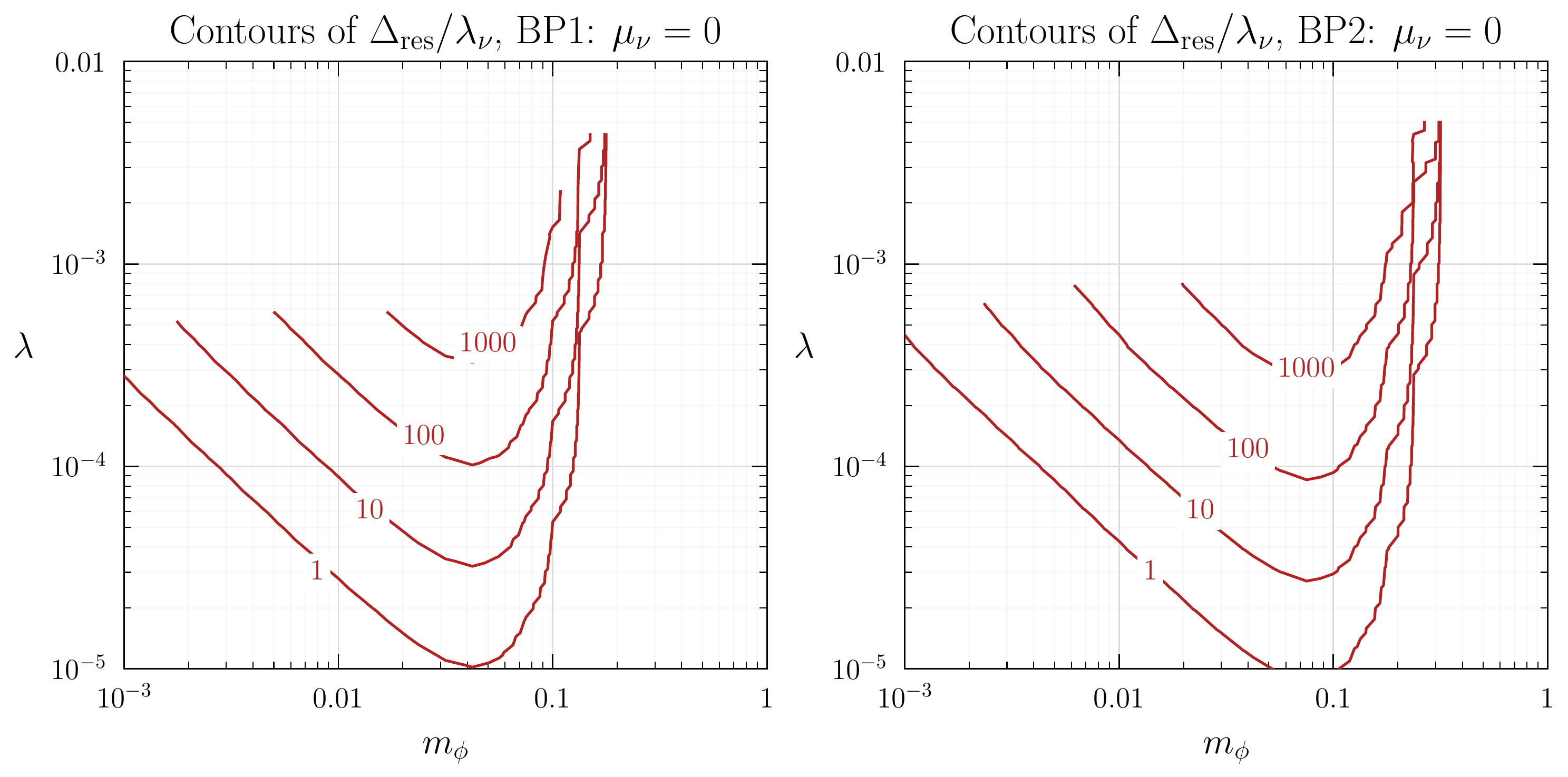}}
	\caption{This figures shows that the ratio $\Delta_{\rm Res}/\lambda_\nu$ is always larger than 1 in the $\lambda$ versus $m_\phi$ plane where we set the SN constraint (in Fig.~\ref{fig:mainresult}). This comparison assures that non-adiabatic transition is unimportant in the region of parameter space of interest to the present work. 
	The contours are shown for benchmark points BP1 and BP2.
	For the case of BP3, the active-sterile neutrino mass difference is so large that resonance does not occur for $E_4=30\,$MeV unless $\lambda$ is sizable $\gtrsim 10^{-3}$. The corresponding $\Delta_{\rm Res}/\lambda_\nu$ is much larger than 1.
	}\label{fig:adiabatic}
\end{figure}

The non-adiabatic transition effect is important only if neutrino passes through the resonance region while it remains free streaming, i.e.,
\begin{equation}\label{eq:adiabatic_condition}
\Delta_{\rm Res} \ll \lambda_\nu \ .
\end{equation}
In order to assess the relevance of this effect, we plot the ratio $\Delta_{\rm Res}/\lambda_\nu$ in the plane of $\lambda$ versus $m_\phi$, as shown in Fig.~\ref{fig:adiabatic}. We first choose a typical SN neutrino energy $E_4=30\,$MeV and solve for the radius $r$ where the resonant condition $\Delta \cos2\theta = V$ is satisfied. We then use Eqs.~\eqref{eq:Delta_res} and \eqref{eq:lambda_nu} to obtain the corresponding $\Delta_{\rm Res}$ and $\lambda_\nu$. As we can see in the figure, the hierarchy in Eq.~\eqref{eq:adiabatic_condition} only occurs for very small values of the coupling $\lambda$. The $\lambda\to0$ limit corresponds to the minimal S$\nu$DM model where neutrino only participates in weak interactions. In that limit, it has been concluded that the feedback effect must be taken into account, which substantially weakens the SN cooling constraint~\cite{Suliga:2019bsq}. 

In the region with $\lambda \gtrsim 10^{-5} - 10^{-4}$, we always find $\Delta_{\rm Res}/\lambda_\nu >1$. Physically, the introduction of neutrino self-scattering mediated by on-shell $\phi-$exchange becomes more frequent in the larger $\lambda$ region which can strongly reduce the neutrino free-streaming length. As a result, the non-adiabatic resonant transition is interrupted by the frequent self-scatterings. The new SN constraints presented in Fig.~\ref{fig:mainresult} and \ref{fig:finalplot} (see below) belong to this region. This confirms that the collisional cooling mechanism taken into account in Sec.~\ref{sec:collisionalcooling} already captures the dominant cooling effect of CCSN in this model.

\section{Discussion and Conclusion}
\label{sec:conclusion}
keV-mass sterile neutrino dark matter (${\rm S\nu DM}$) can be produced efficiently in the early Universe through novel secret neutrino self-interactions. However, the same interactions can also enhance the production of ${\rm S\nu DM}$ in the cores of core-collapse supernovae (CCSN), resulting in faster cooling of the SN. This can lead to a shortened neutrino emission period, which will conflict with observations from SN1987A. This work has computed the constraints on the ${\rm S\nu DM}$ parameter space using SN cooling arguments.

\begin{figure}[!t]
    \centerline{\includegraphics[width=1.0\textwidth]{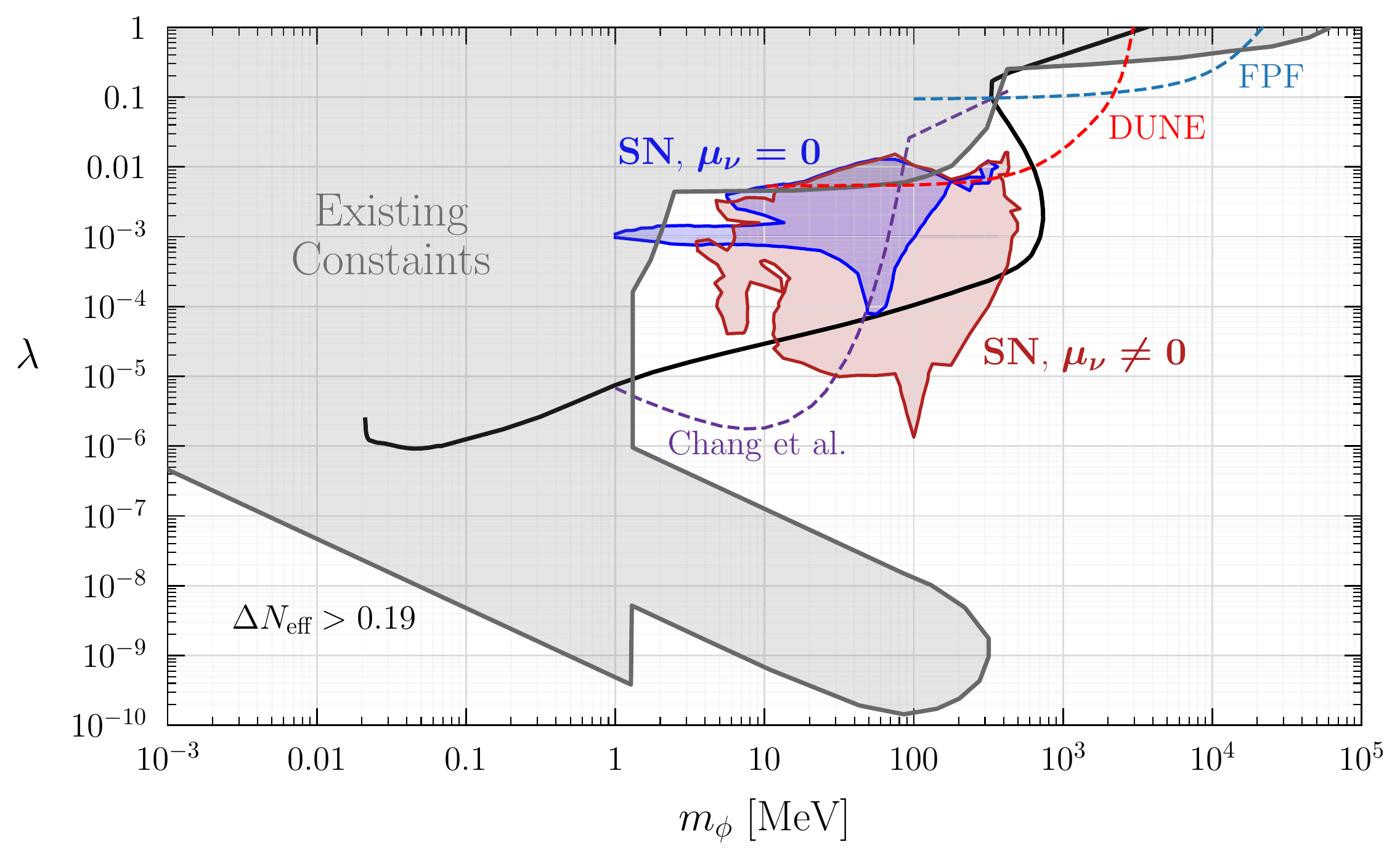}}
	\caption{Summary of various probes of neutrino self-interaction and the S$\nu$DM relic target. Existing constraints and future reaches in the figure are adopted from a recent Snowmass exercise~\cite{Berryman:2022hds} and a very recent work by Chang et al~\cite{Chang:2022aas}. The S$\nu$DM relic target is for BP2 defined in Eq.~\eqref{eq:bps}, with $m_4=21\,$keV and $\sin^22\theta=1.4\times10^{-13}$.
	In the lower-left region ($m_4 < m_\phi\ll 1\,$MeV), the black relic curve can stick out of the gray shaded region labelled as ``Existing Constraints'' (mostly from BBN near the lower edge) only for S$\nu$DM mass below $\sim5$\,keV~\cite{Kelly:2020aks}.
	}\label{fig:finalplot}
\end{figure}

Considering scalar-mediated neutrino self-interactions, we computed in detail the luminosity of ${\rm S\nu DM}$ produced inside the SN core, including the re-scattering effects of the sterile neutrino on its way out. While calculating the thermal contributions to the neutrino scattering, and the potential, we included the presence of a non-zero chemical potential, which is particularly relevant for studying $\nu_e$ self-interactions.  We found that, in the presence of strong neutrino self-interactions, the dominant production mode of ${\rm S\nu DM}$ is through collisional effects. In contrast, production via non-adiabatic transitions can be neglected. A back-of-the-envelope estimate shows that the SN constraints are the strongest for couplings $\sim \mathcal{O}(10^{-3})$, where the thermal potential generated by the new interactions for $\sim \mathcal{O}({\rm MeV})$ mass mediator is comparable to that generated by the weak interactions. Interestingly, as we go to larger couplings, the cooling effect becomes unimportant due to a suppression of the mixing angle in matter. Our results, shown for three benchmark points, indicate that cooling constraints from SN allow us to set novel bounds on the scalar mass-coupling parameter space, which are highly complementary to laboratory and cosmological probes.
We use Fig.~\ref{fig:finalplot} to summarize the interplay among all these frontiers.

One notable assumption in this work is neglecting the dynamical feedback effect on the effective matter potential, due to the production of the sterile neutrinos. This effect, studied in detail in~\cite{Suliga:2019bsq,Suliga:2020vpz} can relax some of the bounds. However, this is most relevant for mixing angles much larger than what this work considers. A detailed treatment feedback effects is beyond the scope of this work. 

In conclusion, keV-mass ${\rm S\nu DM}$, produced through secret neutrino self-interactions, can strongly impact the dynamics of a CCSN. Our findings indicate that a significant part of the model parameter space is susceptible to faster cooling of the SN, thereby reducing the neutrino luminosity. Using this, one can set new SN cooling constraints on the parameter space that can explain the origin of ${\rm S\nu DM}$ relic density. This interplay is more significant for ${\rm S\nu DM}$ heavier than $\sim10\,$keV, and when the active-sterile neutrino mixing and new self-interaction occur through the $\nu_e$ flavor. Our result contributes to the goal of probing the entire novel neutrino self-interaction window for the S$\nu$DM relic target, drawing efforts from astrophysical, cosmological and particle physics experiments.

\section*{Acknowledgements}

We thank Georg Raffelt and Andrea Caputo for useful discussions. Y.C., D.T. and Y.Z. are supported by the Arthur B. McDonald Canadian Astroparticle Physics Research Institute. 
W.T. is supported by the National Science Foundation under Grant No. PHY-2013052. M.S and W.T. would like to express special thanks to the Mainz Institute for Theoretical Physics (MITP) of the Cluster of Excellence PRISMA+ (Project ID 39083149), for its hospitality and support.
W.T. and Y.Z. are grateful to the organizers of the Pollica Summer Workshop supported by the Regione Campania, Università degli Studi di Salerno, Università degli Studi di Napoli “Federico II”, i dipartimenti di Fisica “Ettore Pancini”  and “E R Caianiello”, and Istituto Nazionale di Fisica Nucleare; and to the townspeople of Pollica, IT, for their generosity and hospitality.
This research was also supported by the Munich Institute for Astro, -Particle and BioPhysics (MIAPbP) which is funded by the Deutsche Forschungsgemeinschaft (DFG, German Research Foundation) under Germany´s Excellence Strategy – EXC-2094 – 390783311.

\appendix

\section{$\phi$ mediated thermal potential}\label{app:0}

The exchange of a neutrinophilic scalar $\phi$ at loop level can a generate thermal potential for neutrino and antineutrino.
The contribution comes from the neutrino self-energy diagram shown in Fig.~\ref{fig:SE}. Using the imaginary time formalism, it can be calculated as~\cite{Quiros:1999jp}
\begin{equation}
-i\Sigma(p) = (i\lambda)(i\lambda^*) \int \frac{d^4 k}{(2\pi)^4} i S_F(k) i D_F (p-k) \ ,
\end{equation}
where 
\begin{equation}
\begin{split}
S_F(p) &= \cancel{p} \left\{ \frac{1}{p^2} + 2\pi i \delta(p^2) \left[ \Theta(p^0) n_f^-(p^0) + \Theta(-p^0) n_f^+(p^0)\right] \right\} \ , \\
D_F(p) & = \frac{1}{p^2 - m_\phi^2} - 2\pi i \delta(p^2 - m_\phi^2) 
\left[ \Theta(p^0) n_b^-(p^0) + \Theta(-p^0) n_b^+(p^0)\right] \ ,
\end{split}
\end{equation}
and $\Theta$ is a step function, $\delta$ is a Dirac delta function. We work in the rest frame of the plasma fluid such that
\begin{equation}
n_f^\mp(p^0) = \frac{1}{e^{|p^0|\mp \mu_f}+1} , \quad n_b^\mp (p^0) = \frac{1}{e^{|p^0|\mp \mu_b}- 1} \ .
\end{equation}
Here $f$ stands for neutrino or antineutrino, and $b$ stands for the mediator $\phi$ or $\phi^*$. 
The temperature-dependent part of the self-energy takes the form
\begin{equation}\label{eq:0A3}
\begin{split}
\Sigma(p,T) &= - |\lambda|^2 \int dk^0 \int \frac{d^3\vec{k}}{(2\pi)^3} \left\{ \frac{\cancel{k}}{(k-p)^2 -m_\phi^2} \delta(k^2) \left[ \Theta(k^0) n_f^-(k^0) + \Theta(-k^0) n_f^+(k^0)\right] \right. \\
&+ \left. \frac{\cancel{k}}{k^2} \delta\left((k-p)^2-m_\phi^2\right) 
\left[ \Theta(k^0-p^0) n_b^-(k^0-p^0) + \Theta(-k^0+p^0) n_b^+(k^0-p^0)\right] \right\}\ .
\end{split}
\end{equation}
Using the $\delta$-functions and the on-shell condition of the external neutrino momentum $p^2=0$, we can complete the $k^0$ integral and the angular parts of the $\vec{k}$ integral for each term in Eq.~\eqref{eq:0A3}, 
\begin{align}
\Sigma_{n_f^-} &= \frac{|\lambda|^2}{16\pi^2} \gamma^0 \int dk \left( \frac{m_\phi^2}{2E^2} \log\frac{m_\phi^2}{m_\phi^2+4Ek} + \frac{2k}{E} \right) n_f^-(k) \ , \label{eq:0A4}\\
\Sigma_{n_f^+} &= \frac{|\lambda|^2}{16\pi^2} \gamma^0 \int dk \left( -\frac{m_\phi^2}{2E^2} \log\frac{m_\phi^2}{m_\phi^2-4Ek} + \frac{2k}{E} \right) n_f^+(k) \ , \label{eq:0A5}\\
\Sigma_{n_b^-} &= \frac{|\lambda|^2}{16\pi^2} \gamma^0 \int \frac{kdk}{\epsilon_1} \left( \frac{2k}{E} - \frac{m_\phi^2}{2E^2} \log\frac{m_\phi^2+ 2Ek + 2E \epsilon_1}{m_\phi^2- 2Ek + 2E \epsilon_1} \right) n_b^-(\epsilon_1) \ , \label{eq:0A6}\\
\Sigma_{n_b^+} &= \frac{|\lambda|^2}{16\pi^2} \gamma^0 \int \frac{kdk}{\epsilon_1} \left( \frac{2k}{E} - \frac{m_\phi^2}{2E^2} \log\frac{m_\phi^2+ 2Ek - 2E \epsilon_1}{m_\phi^2- 2Ek - 2E \epsilon_1} \right) n_b^+(\epsilon_1) \ , \label{eq:0A7}
\end{align}
where $k=|\vec{k}|$, $\epsilon_1 =\sqrt{k^2+m_\phi^2}$ and $E$ is the timelike component of the external momentum $p^\mu$. The thermal potential is related to the self-energy as~\cite{Weldon:1982bn}
\begin{equation}
V_\phi =\frac{1}{4E} {\rm Tr} \left( \cancel{p}{\rm Re}\Sigma \right) \ .
\end{equation}
Adding up the pieces of contribution from Eqs.~\eqref{eq:0A4},~\eqref{eq:0A5},~\eqref{eq:0A6},~\eqref{eq:0A7} gives the total $\phi$ mediated potential presented in Eq.~\eqref{eq:V_phi}.

\section{The total luminosity integrals}\label{app:A}

We present, in this section, some details of the derivation of the total cooling luminosity Eq.~\eqref{eq:Lcool2}. Starting from Eq.~\eqref{eq:Lcool}, where the luminosity is originally defined, we first complete the final-state phase-space integrals (F.S.I.) over $\vec{p}_3$ and $\vec{p}_4$. Because of the extra factor $E_4$ in the integrand, the final state integral is not a Lorentz invariant quantity. Here, we have to work in a general reference frame where the center of mass moves with an overall momentum $\vec{P} = \vec{p}_1+\vec{p}_2$. We use the $\delta^4 (\vec{p}_1+\vec{p}_2 - \vec{p}_3 - \vec{p}_4)$ function to reduce the number of integrals down to one,
\begin{equation}
\begin{split}
{\rm F.S.I.} &= \int \frac{d^3 \vec{p}_3}{(2\pi)^3 2E_3} \int \frac{d^3 \vec{p}_4}{(2\pi)^3 2E_4} (2\pi)^4 \delta^4 (\vec{p}_1+\vec{p}_2 - \vec{p}_3 - \vec{p}_4) |\mathcal{M}|^2 E_4 e^{- \tau (E_4, r)} \\
&= \frac{4\pi \lambda^2 m_\phi^2 \delta (s-m_\phi^2)}{|\vec{P}|} \int_{\frac{1}{2}(E_1+E_2 - |\vec{P}|)}^{\frac{1}{2}(E_1+E_2 + |\vec{P}|)} dE_4 \sin^2 \theta_{\rm eff}(r, E_4) E_4 e^{- \tau (E_4, r)} \ ,
\end{split}
\end{equation}
where $|\mathcal{M}|^2$ is given in Eq.~\eqref{eq:M2} and, in the last line,  $s=(p_1+p_2)^2$ is a Mandelstam variable.

Before moving on, we rewrite the remaining $\delta$-function using the kinematic relation
$s=E_1^2 + E_2^2 + 2 E_1 E_2 \cos\theta_{12}$, where $\theta_{12}$ is the relative angle between the three-momenta $\vec{p}_1$ and $\vec{p}_2$. This leads to
\begin{equation}\label{eq:A2}
\delta (s-m_\phi^2) = \frac{1}{2E_1E_2} \delta \left( \cos\theta_{12} - 1 + \frac{m_\phi^2}{2E_1E_2}\right) \ .
\end{equation}
The on-shell condition for the intermediate $\phi$ particle also allows us to derive, $|\vec{P}| = \sqrt{(E_1+E_2)^2 - m_\phi^2}$.

Among the initial state momentum integrals, the integral over the direction of $\vec{p}_1$ is trivial which allows us to pin it along the $\hat z$ axis. The integral over relative angle $\theta_{12}$ can then be completed with the above $\delta$-function, Eq.~\eqref{eq:A2}. This allows us to finally simplify the total luminosity to three energy integrals,
\begin{equation}
\begin{split}
L &= \int d^3 \vec{r} \int \frac{d^3 \vec{p}_1}{(2\pi)^3} f_\nu(E_1, r) \int \frac{d^3 \vec{p}_2}{(2\pi)^3} f_\nu(E_2, r) \frac{1}{4 E_1 E_2} {\rm F.S.I.} \\
&= \frac{\lambda^2 m_\phi^2}{4\pi^2} \int_0^{4 R_c} r^2dr \int_0^\infty dE_1 f(E_1, r) \int_{m_\phi^2/(4E_1)}^\infty dE_2 f(E_2, r) \frac{1}{\sqrt{(E_1+E_2)^2 - m_\phi^2}} \\
&\quad \times \int_{\frac{1}{2}\left( E_1+E_2 - \sqrt{(E_1+E_2)^2 - m_\phi^2} \right)}^{\frac{1}{2}\left( E_1+E_2 + \sqrt{(E_1+E_2)^2 - m_\phi^2} \right)} dE_4 \sin^2 \theta_{\rm eff}(r, E_4) E_4 e^{- \tau (E_4, r)} \ .
\end{split}
\end{equation}
This is Eq.~\eqref{eq:Lcool2} that we use to proceed with the numerical analysis.

\section{Width of MSW resonance inside supernova}\label{app:B}

The Hamiltonian governing active-sterile neutrino oscillation in the SN takes the form
\begin{equation}
H = \begin{pmatrix}
\Delta \sin^2 \theta + V(r) & \Delta \sin\theta \cos\theta \\
\Delta \sin\theta \cos\theta & \Delta \cos^2 \theta
\end{pmatrix} \ .
\end{equation}
It can be diagonalized with a unitary transformation
\begin{equation}
H = 
\begin{pmatrix}
\cos\theta_m & \sin \theta_m \\
-\sin\theta_m & \cos\theta_m
\end{pmatrix}
\begin{pmatrix}
E_1 & 0 \\
0 & E_2
\end{pmatrix}
\begin{pmatrix}
\cos\theta_m & -\sin \theta_m \\
\sin\theta_m & \cos\theta_m
\end{pmatrix} \equiv U^\dag \hat H U\ ,
\end{equation}
where $E_1, E_2$ and $\theta_m$ are functions of $r$, and satisfy
\begin{align}
\tan^2 2\theta_m(r) &= \frac{\Delta^2 \sin^22\theta}{(\Delta \cos2\theta - V(r))^2} \ , \label{eq:B3}\\
\Delta E(r) = |E_1-E_2| &= \sqrt{\Delta^2\sin^22\theta + (\Delta \cos2\theta - V)^2} \ .
\end{align}

We first consider an ultra-relativistic neutrino propagating in the radial direction of the supernova, with $dr = c dt$. The Schr\"odinger equation takes the form
\begin{equation}
i \hbar c \frac{\partial}{\partial r} \psi = H \psi = U^\dagger \hat H U \psi \ .
\end{equation}
Defining $\psi_m=U\psi$, the equation for $\psi_m$ is
\begin{equation}
i \hbar c \frac{\partial}{\partial r} \psi_m = \hat H \psi_m - i \hbar c \left( U \frac{\partial U^\dagger}{\partial r} \right) \psi_m \ .
\end{equation}
In this new basis, although $\hat H$ is a diagonal matrix, the additional term proportional to $U \partial U^\dagger/\partial r$ contributes an off-diagonal element
\begin{equation}
\Delta \hat H_{12} = -i \hbar c \left( U \frac{\partial U^\dagger}{\partial r} \right)_{12} = i \hbar c \theta_m' \ .
\end{equation}
Hereafter ${}^\prime$ means $\partial / \partial r$.

Using Eq.~\eqref{eq:B3}, we obtain
\begin{equation}
\theta_m'= \frac{1}{2} \sin2\theta_m \cos2\theta_m \frac{V'}{\Delta \cos2\theta - V} \ .
\end{equation}
Near the MSW resonance, we have 
\begin{align}
\theta_m &\simeq \pi/4 \ , \\
\sin2\theta_m &\simeq 1 \ , \\
\cos2\theta_m &\simeq \frac{\Delta\cos2\theta - V}{\Delta \sin2\theta} \to 0 \ .
\end{align}
These lead to
\begin{equation}
\theta_m' \simeq \frac{V'}{2\Delta \sin2\theta} \simeq \frac{1}{2\tan2\theta} \frac{V'}{V} \ .
\end{equation}

The width of resonance is introduced to characterize the importance of $\Delta \hat H_{12}$,
\begin{equation}
\Delta_{\rm Res} \equiv \frac{\hbar c}{|\Delta \hat H_{12}|} = \frac{1}{|\theta_m'|} = 2 \tan2\theta \left| \frac{V'}{V} \right|^{-1} \ .
\end{equation}

The regular adiabatic condition corresponds to $|\Delta \hat H_{12}| \ll \Delta E(r)$. In the presence of frequent hard scattering among active neutrinos, non-adiabatic transition is also negligible if $\Delta_{\rm Res} \gg \lambda_\nu$, where $\lambda_\nu = 1/\Gamma$ is the neutrino free streaming length. 

For neutrino traveling along non-radial directions, the variations of background temperature and densities in the CCSN are slower, thus $|\theta_m'|$ is reduced, leading to a larger $\Delta_{\rm Res}$. In that case, it would be safer to neglect non-adiabatic transitions.

\bibliography{references}
\end{document}